\documentclass[pre,twocolumn,amsmath,amssymb,superscriptaddress,showpacs]{revtex4-1}
\usepackage{graphicx}
\usepackage{latexsym}
\usepackage{dcolumn}
\usepackage{bm}
\usepackage{color}
\usepackage{amssymb}
\usepackage[normalem]{ulem}

\newcommand{\be}{\begin{equation}}
\newcommand{\ee}{\end{equation}}
\newcommand{\bea}{\begin{eqnarray}}
\newcommand{\eea}{\end{eqnarray}}
\newcommand{\bse}{\begin{subequations}}
\newcommand{\ese}{\end{subequations}}

\newcommand{\e}{\epsilon}

\newcommand{\comment}[1]{}

\begin{document}

\title{Finite-size localization scenarios in condensation transitions}
\date{\today}

\author{Gabriele Gotti}
\affiliation{Dipartimento di Fisica e Astronomia, Universit\`a di Firenze,
via G. Sansone 1 I-50019, Sesto Fiorentino, Italy}
\affiliation{Istituto dei Sistemi Complessi, Consiglio Nazionale
delle Ricerche, via Madonna del Piano 10, I-50019 Sesto Fiorentino, Italy}

\author{Stefano Iubini}
\affiliation{Istituto dei Sistemi Complessi, Consiglio Nazionale
delle Ricerche, via Madonna del Piano 10, I-50019 Sesto Fiorentino, Italy}
\affiliation{Istituto Nazionale di Fisica Nucleare, Sezione di Firenze, 
via G. Sansone 1 I-50019, Sesto Fiorentino, Italy}

\author{Paolo Politi}
\email{paolo.politi@cnr.it}
\affiliation{Istituto dei Sistemi Complessi, Consiglio Nazionale
delle Ricerche, via Madonna del Piano 10, I-50019 Sesto Fiorentino, Italy}
\affiliation{Istituto Nazionale di Fisica Nucleare, Sezione di Firenze, 
via G. Sansone 1 I-50019, Sesto Fiorentino, Italy}

\begin{abstract}
We consider the phenomenon of condensation of a globally conserved
quantity $H=\sum_{i=1}^N \e_i$ distributed on $N$ sites, 
occurring when the density $h= H/N$ exceeds a critical density $h_c$.
We numerically study the dependence of the participation ratio $Y_2=\langle \epsilon_i^2\rangle/(Nh^2)$ 
on the size $N$ of the system and on the control parameter $\delta = (h-h_c)$,
for various models: (i)~a model with two conservation laws, derived from the Discrete NonLinear Schr\"odinger
equation; (ii)~the continuous version of the Zero Range Process class, 
for different forms of the function $f(\e)$ defining the factorized steady state.
Our results show that various localization scenarios may appear for finite $N$ and close to the transition point.
These scenarios are characterized by the presence or the absence of a minimum of
$Y_2$ when plotted against $N$ and by an exponent $\gamma\geq 2$ 
defined through the relation $N^* \simeq \delta^{-\gamma}$, where $N^*$
separates the delocalized
region ($N\ll N^*$, $Y_2$ vanishes with increasing $N$) from the localized region
($N\gg N^*$, $Y_2$ is approximately constant).
We finally compare our results with the structure of the condensate obtained
through the single-site marginal distribution.
\end{abstract}
\maketitle

\section{Introduction} 
\label{sec.intro}
The word {\it condensation} typically refers to the condensation of a gas~\cite{book_Chaikin},
a classical change of the physical state of matter, or to
the Bose-Einstein condensation (BEC),
a quantum phenomenon observed in dilute atomic gases~\cite{book_Huang}. 
In the last twenty years it has acquired a new meaning related
to transport phenomena:
under certain conditions it may appear a critical density of the transported quantity
(e.g., mass) above which
a finite fraction of whole mass condenses/localizes at a single site of a lattice~%
\cite{Drouffe1998_JPA,Majumdar2005_PRL,Godreche2007_LNP,Majumdar2010_LesHouches,Zannetti2014_PRE,Godreche2020_arXiv}.
It is therefore a localization in the real space while BEC is a localization in the momentum space.

The phenomenon of real-space condensation appears ubiquitously in several domains of physics,
ranging from shaken granular materials~\cite{eggers99,torok05} to the behavior of traffic flows~\cite{Evans2005_JPA}, just to mention a couple of
important applications. 
Another  physically relevant example appears when studying 
the Discrete NonLinear Schr\"odinger (DNLS) equation~\cite{kevrekidis09},
which is used to model nonlinear wave propagation in weakly dissipative systems with important applications
in the fields of nonlinear optics~\cite{ESMBA} and cold atoms~\cite{TS}.
In this context, in spite of the presence of two conservation laws (see Sec.~\ref{sec.C2C}),
it has recently been shown~\cite{Gradenigo2021_JSTAT,Gradenigo2021_EPJE}
that the condensation transition of the DNLS model is strictly related to 
a special case of the class of zero-range processes (ZRP)~\cite{Evans2005_JPA},
where only one quantity is conserved.
In the wide class of ZRP, the sites of a lattice hosts indistinguishable particles which hop from a site $i$ to a neighboring site
$i'$ with a rate depending only on the number of particles at the departure site $i$. 
The peculiarity of ZRP class and of its continuous version~\cite{evans04jpa,zia04} 
is that stationary/equilibrium states are given by a simple factorized form~\cite{Evans2005_JPA}, which  may allow for analytic descriptions of the condensation process~\cite{Evans2006_JSP}.
	

Finite-size effects play a  relevant role in condensation 
transitions, as they can determine the dominant behavior of the system even for moderately large system
sizes. As an example, in the DNLS model they are responsible of the emergence of 
peculiar delocalized states at negative absolute temperature. Such states disappear in the thermodynamic limit, but they are 
still easily detectable even
for lattices of thousands of sites~\cite{Gradenigo2021_JSTAT,Gradenigo2021_EPJE}. 
 %
 
 In this paper we provide a systematic numerical study of the
localization process occurring at finite sizes and for different classes of condensation models.   
%
%
%
%
In order to be specific,
\comment{During the years, attention has been paid both to nonequilibrium and equilibrium properties. Here we limit ourselves to
consider the equilibrium ones and we mention two important results
about the condition because condensation appears and about 
the nature of the condensate~\cite{Majumdar2005_PRL,Evans2006_JSP}.
These results do not depend on the discrete nature of the transported and globally conserved quantity
and can be generalized to ``continuous mass" models~\cite{evans04jpa,zia04} which are
more appropriate for the present work. 
}
let us suppose to have $N$ independent random variables $\e_i$, equally distributed
according to some distribution function $f(\e)$ and constrained by their sum,
$\sum_i \e_i = H \equiv Nh$~\cite{Godreche2001_EPJB}.
Condensation is known to occur if asymptotically ($\e\to\infty$)~\cite{Evans2006_JSP}
\be
\exp(-\e) < f(\e) < \frac{1}{\e^2} ,
\label{eq.condcond}
\ee
and if $h>h_c$, with the critical value for condensation given by 
\be
h_c = \int_0^\infty \!\!d\e\, \e f(\e) \; . 
\ee

\comment{The probability of finding the system in the configuration $\{\epsilon_k\}=(\epsilon_1,\cdots,\epsilon_N)$
is equal to
\be
P(\{\epsilon_k\})=\frac{\prod_{k=1}^N f(\e_k)}{Z_N(E)} \delta(E-\sum_i \e_i)
\label{eq.P}
\ee
where the partition function $Z_N(E)$ is the normalization factor.
Accordingly, the single-site marginal distribution of energy $p(\e_i)$ on a single site $i$ is 
obtained by integrating $P(\{\epsilon_k\})$ over all $k\ne i$,
}
In the condensed region, $h\ge h_c$, and in the thermodynamic limit $N\to\infty$, 
the equilibrium state is trivial: one single site $i^*$ collects the ``extra" energy
above the threshold, $\e_{i^*}  = \delta N \equiv (h-h_c)N$, while the ``critical"
energy $h_c N$ is distributed to all other sites according to $f(\e)$.

When the system is finite, the above picture must be corrected evaluating the single-site
marginal distribution,
\be
p(\e) = f(\e) \frac{Z_{N-1}(E-\e)}{Z_N(E)},
\label{eq.dressed}
\ee
where 
$Z_N(E) = \int d\e_1\dots d\e_N \prod_{k=1}^N f(\e_k) \delta (E-\sum_i \e_i)$
is the partition function.
This has been done in Ref.~\cite{Evans2006_JSP} where it is derived the limiting shape
of the condensate, i.e. the form of the distribution (\ref{eq.dressed}) around $\e =\delta N$.

Here we focus on the finite-size effects of the condensation transition close
to the critical point ($\delta\ll 1, N\gg 1$) using the participation ratio,
\be
Y_2(N,\delta) \equiv \frac{\langle\, \overline{\e_i^2}\,\rangle}{\displaystyle N\langle\,\overline{\e_i}\,\rangle^2} =
\frac{\langle \,\overline{\e_i^2}\,\rangle}{Nh^2} \,,
\label{eq.Y2}
\ee
where $\overline{(\cdots )}$ is a spatial average and $\langle\cdots\rangle$ is a statistical average.
$Y_2$, firstly introduced in the theory of electron localization~\cite{Thouless1974_review}, 
has been recently used in the context of condensation transitions~\cite{Gradenigo2017_Entropy,Gradenigo2021_JSTAT,Gradenigo2021_EPJE}.
Most importantly, it should be noted that $Y_2$ is a collective property which
represents the proper order parameter for the condensation/localization
transition,  as it allows to discriminate between homogeneous and localized states. 
In the former case $\langle\,\overline{\e_i}\,\rangle^2$ is finite and 
$Y_2$ asymptotically vanishes as $1/N$.
In the latter case the average is dominated by the single site hosting the extra-energy, 
$\langle \,\overline{\e_i^2}\,\rangle = N\delta^2 + o(N)$  
and  in the thermodynamic limit $Y_2$ remains finite:
\be
Y_2^\infty (\delta) = \frac{\delta^2}{h^2} = \frac{\delta^2}{(h_c +\delta)^2} \,.
\ee

For finite system sizes it is no more possible to assume that a single site hosts
all the excess energy as soon as $\delta >0$ and the evaluation of $Y_2$
is much more complicated.
The full knowledge of $p(\e)$, see Eq.~(\ref{eq.dressed}), would allow to compute $Y_2(N,\delta)$
but in practice this is hardly feasible in general.
For this reason in the following we will carry out a detailed numerical analysis of 
how $Y_2(N,\delta)$ depends on the size $N$ 
and on the reduced control parameter $\delta= h-h_c$, for several models.
In particular, the use of the participation ratio will allow to characterize the condensation transition through
some qualitative and quantitative features which will be seen to depend on the asymptotic form of $f(\e)$.
A detailed description of the content of the paper follows.

We  start in Sec.~\ref{sec.C2C} with the DNLS equation, showing the well known
result that close to the transition point it reduces to a simple microcanonical model with
two conservation laws, called here C2C model.
Our analysis shows that $Y_2(N,\delta)$ does not tend to the asymptotic value $Y_2^\infty (\delta)$
monotonically: it displays a minimum
as a function of $N$ and $Y_2^{min}(\delta) \sim\delta^\gamma$ with $\gamma=3$.
The study of the participation ratio therefore allows to introduce the exponent $\gamma$.

\comment{
More recent works have stressed the importance of having more than one conservation law,
both in terms of the chance to have an otherwise-absent condensation transition~%
\cite{Iubini2013_NJP,JSP_DNLS,JSTAT_mmc,Szavits2014_PRL,Szavits2014_JPA}
and in terms of formal connections between different models~\cite{Gradenigo2021_JSTAT}.
}

In Sec.~\ref{sec.ZRP} we pass to study different cases of the ``continuous mass"
transport models, i.e. different functions $f(\e)$ in the condensation range
defined in Eq.~(\ref{eq.condcond}).
Since the transported quantity is continuous and we focus on the equilibrium properties 
we will use the acronym FCT (Factorized Continuous Transport) to indicate these models.
We will start by discussing a recent result~\cite{Gradenigo2021_JSTAT} 
according to which the C2C model is asymptotically equivalent to the FCT model
where $f(\e)$ is a stretched-exponential distribution, $f(\e)\sim \exp(-\sqrt\e)$.
The following analysis of the participation ratio for the cases where $f(\e)$ decays as a power law $1/\e^\beta$
shows localization scenarios which are different from the C2C one: 
when $\beta=4$, $Y_2$ still displays a minimum as a function of $N$ but with
a different exponent $\gamma=2$, because now $Y_2^{min}(\delta) \sim\delta^2$.
Moreover, when $\beta=2.5$ there is no minimum at all.
All the scenarios emerging from numerics will be interpreted in terms of the effective number of sites
hosting the condensate, $K(N,\delta)$.

Finally, in Sec.~\ref{sec.discussion} we summarize our results and
discuss the different localization scenarios in the light of related studies 
on ZRP/FCT models~\cite{Evans2006_JSP}
and on the C2C model~\cite{Gradenigo2021_JSTAT}.

\section{The microcanonical DNLS/C2C model}
\label{sec.C2C}

The one-dimensional Discrete NonLinear Schr\"odinger (DNLS) model is
defined by the Hamiltonian~\cite{kevrekidis09} 
\begin{equation}
{ H}= \sum_n \left( |z_n|^4+z_n^*z_{n+1}+z_nz_{n+1}^* \right) ,
\label{Hz}
\end{equation}
where $z_n = \rho_n e^{i\phi_n}$ are complex variables whose canonically-conjugated variables
are $-iz^*_n$. Accordingly, the DNLS equation is derived from the Hamilton's equations $\dot z_n = \partial { H}/\partial (-iz^*_n)$ and reads,
\begin{equation}
i\dot{z}_n = -2|z_n|^2z_n -z_{n+1} -z_{n-1}\,.
\end{equation}

This model has two conserved quantities, the energy (because ot time invariance of 
${ H}$) and the mass $A=\sum_n |z_n|^2$ (because of the invariance under global phase
transformations, $z_n \to z_n e^{i\bar\phi}$). Two results are now well established~%
\cite{Rasmussen2000_PRL,Iubini2013_NJP,JSP_DNLS,Szavits2014_PRL, JSTAT_mmc,Gradenigo2021_JSTAT}:
(i)~this model has an equilibrium localized state for $h>h_c=2a^2$, where $a=A/N$ is the mass
density, ${ h}={ H}/N$ is the energy density, and the curve $h_c=2a^2$ also corresponds
to the infinite-temperature line.
(ii)~Close to this line the coupling terms between neighboring sites (see Eq.~(\ref{Hz})), $z_n^*z_{n+1}+ z_nz_{n+1}^*$,
are negligible~\cite{Gradenigo2021_JSTAT}, so that the phases $\phi_n$ can be
ignored. 
In this limit the DNLS model is equivalent to a much simpler system,
which is defined in terms of the positive-definite local masses $c_i = \rho_i^2$  and their squares (local energies) $\e_i = c_i^2$. 
The total mass and energy
\be
A = \sum_{i=1}^N c_i, \qquad H = \sum_{i=1}^N \e_i.
\ee
are conserved
and it can be shown~\cite{Gradenigo2021_JSTAT} that the condensation condition  $h>h_c=2a^2$
still holds, with the same meaning of the symbols $a$ and $h$ defined above.
We refer to this model as the C2C model. In this model
the density $a$ is just a unit of measure of the mass and we can
set $a=1$ without loss of generality.

The existence of a condensation transition can be understood in the C2C model as follows.
The ground state corresponds to
a perfectly homogeneous state with
all sites hosting the same mass, $c_i\equiv 1$, so $h_{GS}=1$.
Upon increasing the energy ($h > 1$), mass fluctuations increase evenly 
until $h=2$, where $\langle c_i^2\rangle = 2\langle c_i\rangle^2$.
Therefore at $h=2$ the standard deviation
of the mass distribution is equal to its average value and
the injection of additional energy is expected to be localized.
According to~\cite{Rasmussen2000_PRL} at criticality ($h=2$) the mass is distributed
exponentially
\be
\tilde f(c)= \exp(-c) .
\label{eq.exp}
\ee
Above criticality (and in the thermodynamic limit!) 
a single site hosts the extra energy $\delta N$.

\begin{figure}
\begin{center}
\includegraphics[width=0.45\textwidth,clip]{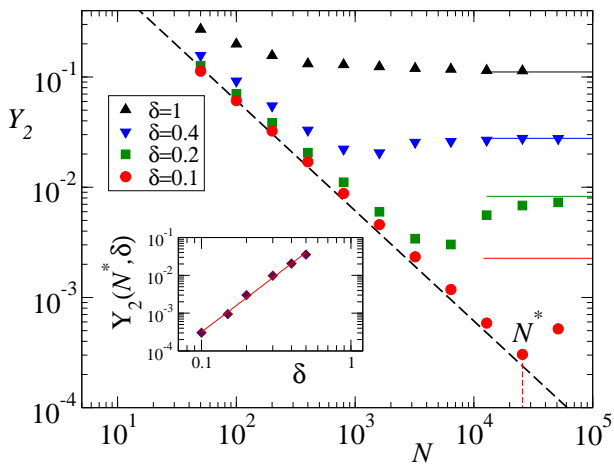}
\end{center}
\caption{
Curves of $Y_2(N,\delta)$ of the C2C model for some values of $\delta$, see legend.
The black dashed curve shows the analytical scaling $Y_2=6/N$, expected for $N \ll N^*$.
Full lines are the analytical asymptotic values, equal to $\delta^2/h^2$. 
Inset: 
behavior of the minimum $Y_2(N^*,\delta)$ as a function of $\delta$ (black diamonds). 
The red line shows a growth as $\delta^3$.
}
\label{fig.C2CY2}
\end{figure}

The equilibrium properties of the C2C model, in particular the evaluation of the participation ratio 
$Y_2(N,\delta)$, see Eq.~(\ref{eq.Y2}), have been obtained here numerically through a Microcanonical Monte Carlo algorithm.
The simplest stochastic evolution rule satisfying detailed balance and conservation laws
is~\cite{JSP_DNLS} to choose a triplet of distinct sites $(i,j,k)$ and update 
their masses in order to conserve their sum
(the mass) and the sum of their squares (the energy). 
This procedure amounts to determine the intersection between the plane
$c_i+c_j+c_k = M$ with the sphere $c_i^2 + c_j^2 +c_k^2 = E$ and to choose a random point in such intersection.
If the masses are of the same order the intersection is a circle, otherwise the positivity
constraint on the masses limits the intersection to the union of three arcs.
The three sites can be chosen either in sequence, ($i-1,i,i+1$), or randomly.
This choice affects the relaxation dynamics towards the equilibrium condensed state,
which displays a coarsening process and
has been studied in detail~\cite{JSP_DNLS,JSTAT_mmc} for a triplet of neighboring sites.
Here, instead, we will focus on the properties of the equilibrium state,
which are not affected by such a choice. Accordingly, we implement the update of random triplets of sites
in order to speed up numerical simulations~\footnote{We have always verified for all numerical simulations 
that the equilibrium state was reached: suitable relaxation transients were introduced to
make the system relax before computing statistical averages.}.
Finally, we stress that the energy $h$ does not enter explicitly  the evolution rule: 
$h$ only specifies the initial configuration, which is appropriately chosen so as
to have $\bar \e_i =h$.

In Fig.~\ref{fig.C2CY2} we plot $Y_2$ versus $N$ for different values of $\delta$. 
The main feature of these curves is that $Y_2$ does not tend to the asymptotic value
$Y_2^\infty (\delta) =\delta^2/h^2$ monotonically: 
except for large values of $\delta$, the participation ratio has a minimum,
occurring for $N=N^*(\delta)$ and separating the delocalized regime (where $Y_2$ decays as $1/N$ and does not depend
on $\delta$) from a localized one, where $Y_2 \to Y_2^\infty(\delta)$. 
Furthermore, according to the inset the participation ratio at the minimum 
vanishes with $\delta$ as a power law, $Y_2(N^*,\delta) \simeq \delta^\gamma$,
which allows to define the exponent  $\gamma =3$.

\begin{figure}
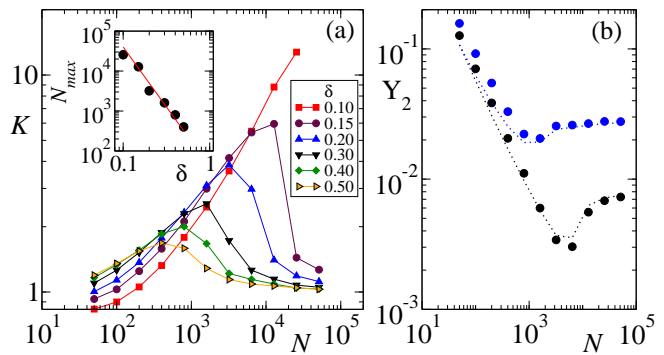

\begin{center}
\includegraphics[height=4.6 cm,clip]{C2CK}
\includegraphics[height=4.6 cm,clip]{C2CY2num-theo}
\end{center}
\caption{
(a)~Effective number of localization sites for the C2C model, $K(N,\delta)$,
for different values of $\delta$.
Inset: 
Size $N_{max}$ for which the maximum of $K(N)$ is attained versus $\delta$. The red curve is 
a power-law scaling as $\delta^{-3}$.
(b)~
Participation ratio directly obtained from numerics (full circles) and reproduced from Eq.~(\ref{eq.Y2anal})
using data for $K$ as given in (a)
(dotted lines). Data refer to $\delta=0.4$ (blue symbols, upper curve) and $\delta=0.2$ (black symbols, lower curve).
}
\label{fig.C2Cvarie}
\end{figure}

In the absence of an exact analytic expression for $Y_2(N,\delta)$
we propose a phenomenological interpretation of our results which is also applicable to 
the models studied in the next Section.
The basic assumption is that the extra energy $\delta N$ is equally shared among $K(N,\delta)$ sites, while
the remaining energy is homogeneously distributed on the other sites with a second moment equal to $\mu_2$. 
With these hypotheses from Eq.~(\ref{eq.Y2}) we obtain
\be
Y_2(N,\delta) = \frac{\mu_2}{Nh^2} + \frac{\delta^2}{K(N,\delta)h^2} + o\left(\frac{1}{N}\right) \, ,
\label{eq.Y2anal}
\ee
which immediately clarifies that the condition $K=1$ (or $K$ any constant)  cannot produce a minimum in $Y_2$. 
More precisely, if we take the derivative of $Y_2$ with respect to $N$ we obtain
\be
\frac{dY_2}{dN} = -\frac{\mu_2}{N^2h^2} - \frac{\delta^2}{K^2h^2}\frac{dK}{dN}, 
\ee
indicating that a minimum of $Y_2(N,\delta)$ ($dY_2/dN=0$) may appear only if $K(N,\delta)$
decreases in some interval of $N$ ($dK/dN <0$).

The correctness of this interpretation is 
shown in Fig.~\ref{fig.C2Cvarie}. 
In left panel (a) the effective number of sites hosting the extra energy, $K(N,\delta)$,
has been {\it independently} derived from simulations
comparing the extra energy with the actual, maximal energy appearing in the lattice:
their ratio gives an estimate of $K(N,\delta)$.
The curves $K(N,\delta)$ have a maximum which can explain the minimum of the
participation ratio. Furthermore, the position of the  maximum $N_{max}$ moves to larger $N$ when 
$\delta$ decreases (which is in agreement with a decreasing function $N^*(\delta)$)
and the maximum itself increases upon decreasing $\delta$
(which is in agreement with the fact that $Y_2(N^*)$ decreases more rapidly than
$Y_2^\infty$). 
The inset of Fig.~\ref{fig.C2Cvarie}(a) shows that $N_{max}(\delta)$
scales as $\delta^{-3}$, that is, as $N^*(\delta)$.
If we now make use of the just determined function $K(N,\delta)$ in Eq.~(\ref{eq.Y2anal}),
we obtain the curves $Y_2(N,\delta)$ as given in Fig.~\ref{fig.C2Cvarie}(b) (see dotted lines), 
which satisfactorily reproduce the behavior of the participation ratio in the whole range of $N$,
attesting the correctness of the proposed expression (\ref{eq.Y2anal}) for $Y_2(N,\delta)$.

The next step is to understand whether the features of $Y_2(N,\delta)$ we have found for the C2C model
(existence of a minimum and value of the exponent $\gamma$) are
either model-specific or generic. For this reason we now move to study the more
general class of FCT models.
This passage is even more justified by the strict connection between C2C and FCT, as now discussed.

\section{Canonical FCT models}
\label{sec.ZRP}

In Ref.~\cite{Gradenigo2021_JSTAT}  the microcanonical
partiton function of the C2C model, 
\be
\Omega(A,E) = \int_0^\infty\!\! \prod_i dc_i\,\, \delta\!\left(A -\sum_i c_i\right)
\delta\!\left(E -\sum_i c_i^2\right) , 
\ee
was studied by means of large-deviations techniques. It was shown that the constraint on the mass 
conservation can  be relaxed by 
taking the Laplace transform
with respect to $A$, explicitly
\be 
\tilde\Omega(\lambda,E) = \left( \frac{1}{2\lambda}\right)
\int_0^\infty\!\! \prod_i d\e_i f_\lambda(\e_i) \delta\!\left(E -\sum_i \e_i\right) 
\label{eq.LTC2C}
\ee
where~\footnote{Eq. (\ref{eq.C2Cstretched})
corresponds to a Weibull distribution~\cite{Weibull} in the variable $\e$ with shape parameter $1/2$ and scale parameter $\lambda^{-2}$.} 
\be
f_\lambda(\e) =\frac{\lambda}{2\sqrt{\e}} \exp (-\lambda\sqrt{\e}).
\label{eq.C2Cstretched}
\ee

Equation (\ref{eq.LTC2C}) can be therefore interpreted as the probability distribution of
the sum of $N$ i.i.d. random variables $\e_i$ distributed according to $f_\lambda(\e)$
and constrained by the their sum.
Since $f_\lambda(\e)$ satisfies the bounds of Eq.~(\ref{eq.condcond}) this is a
further proof that the C2C model has a condensation transition.
The equivalence between the microcanonical C2C model and the canonical FCT model
with a distribution given by Eq.~(\ref{eq.C2Cstretched}) is only guaranteed in the
thermodynamical limit. For this reason we have explicitly simulated the canonical
FCT model with a pure stretched exponential,
\be
f_s(\e) = \frac{3}{2}\exp (-\sqrt{3\e}),
\label{eq.stretched}
\ee
where the $\sqrt{3}$ factor at the exponent has been chosen so as to have 
$h_c= \int_0^\infty \!\! d\e\, \e f_s(\e) = 2$.

\begin{figure}
\begin{center}
\includegraphics[width=0.48\textwidth,clip]{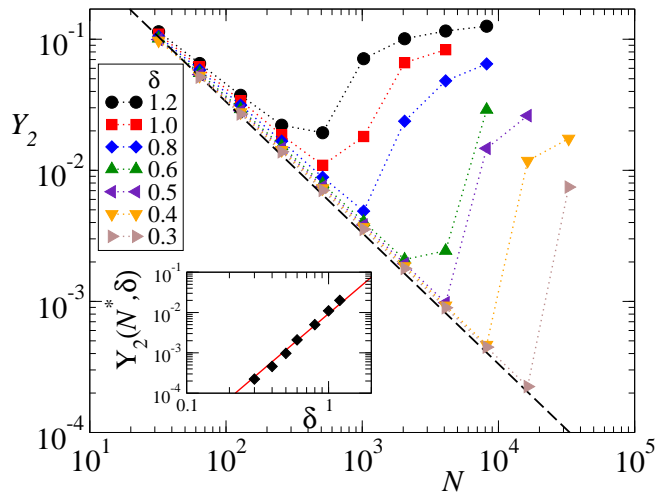}
\end{center}
\caption{
Curves of $Y_2(N,\delta)$ for $f_s(\e) = \frac{3}{2}\exp (-\sqrt{3\e})$.
The black dashed curve shows the analytical scaling $Y_2=10/(3N)$, expected for $N \ll N^*$. 
Inset:  behavior of the minimum $Y_2(N^*,\delta)$ as a function of $\delta$ (black diamonds). 
The red line shows a growth as $\delta^3$.
}
\label{fig.stretched}
\end{figure}

Simulations have been performed using the following Monte Carlo algorithm
to sample the equilibrium state~\cite{Godreche2001_EPJB}.
Two distinct sites $i,j$ are randomly chosen and a random energy value $\e^*$ is extracted from a uniform distribution 
in the interval $(0,\e_i +\e_j)$. The attempted new energies
are $\e'_i = \e^*$ and $\e'_j = (\e_i + \e_j) - \e^*$. This move, which by construction conserves the energy,
 is accepted with probability 
\bea
q &=& \min \{ 1, q^* \}, \nonumber \\
q^* &=& \frac{f_\beta(\e'_i) f_\beta(\e'_j)}{f_\beta(\e_i) f_\beta(\e_j)},
\label{eq.algiid}
\eea
which satisfies detailed balance.
Results are shown in Fig.~\ref{fig.stretched}: not only a clear minimum of the participation ratio appears,
it also scales with an exponent compatible with $\gamma=3$.

In order to check how peculiar are the properties of the C2C model and of the FCT model with
$f(\e) = f_s(\e)$, 
it is now necessary to study definitely different $f(\e)$.
For this reason we will focus to the family of power-law distributions
\be
f_\beta (\e) = \frac{1}{b_0 + \e^\beta},
\label{eq.fbeta}
\ee
where the real parameter $\beta$ satisfies the condition $\beta > 2$ in order to have a condensation phenomenon,
see Eq.~(\ref{eq.condcond}), and $b_0$ is fixed to have the same critical value as the C2C model,
\be
h_c = \frac{\int_0^\infty d\e \e f_\beta (\e)}{\int_0^\infty d\e f_\beta (\e)} = 2 .
\ee

Expression (\ref{eq.Y2anal}) for the participation ratio suggests that a distinction should be made between $\beta >3$
and $\beta \le 3$. In the former case $f_\beta(\e)$ has a finite second moment, while in the latter
case $\mu_2$ diverges as $N^{3-\beta}$ (logarithmically for $\beta=3$)%
~\footnote{This is because for a system of size $N$ the maximal possible energy on a site is $Nh$, not infinity.}. 
All the same, the first term on the RHS of Eq.~(\ref{eq.Y2anal})
 vanishes asymptotically for $\beta >2 $. Therefore in such limit 
it is negligible with respect to the second term which tends to $\delta^2/h^2$.

\begin{figure}
\begin{center}
\includegraphics[width=0.45\textwidth,clip]{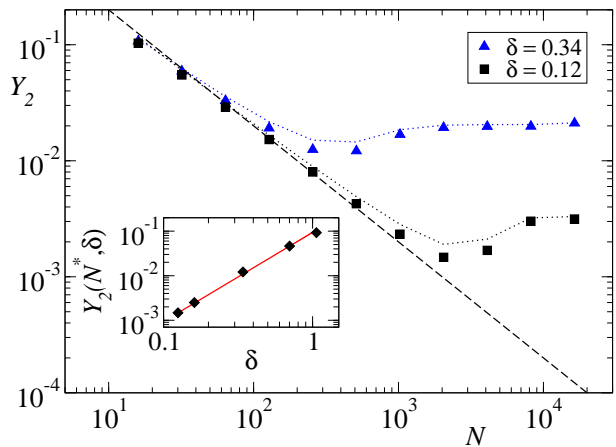}
\end{center}
\caption{
Participation ratio of the $\beta=4$ model for a couple of values of $\delta$ (full symbols).
Dotted lines refer to the values obtained from Eq.~(\ref{eq.Y2anal}). 
The straight dashed line is the analytical expression $Y_2(N,\delta=0)=2/N$, valid for $N\ll N^*$.
Inset: $Y_2(N^*,\delta)$ as a function of $\delta$ (black diamonds). The red line shows a growth as $\delta^2$.
}
\label{fig.beta4}
\end{figure}

\begin{figure}
\begin{center}
\includegraphics[width=0.45\textwidth,clip]{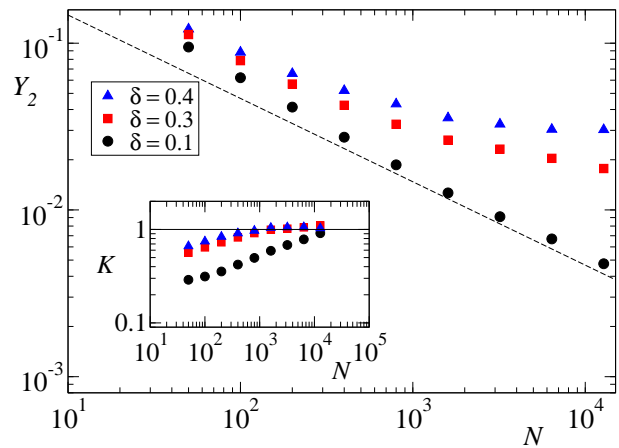}
\end{center}
\caption{
Participation ratio of the $\beta=2.5$ model for different values of $\delta$. 
$Y_2(N,\delta)$ curves at $\delta$ fixed have no minimum, in agreement with the absence of a maximum 
in the curves $K(N)$, see the inset.
Furthermore, for $N\ll N^*$ $Y_2(N)$ decays as $1/N^{1/2}$, see dashed line. 
}
\label{fig.beta2.5}
\end{figure}

We have studied this class of models for $\beta=4$ and $\beta=2.5$ 
Results are shown in Fig.~\ref{fig.beta4} for $\beta=4$ and in Fig.~\ref{fig.beta2.5} for $\beta=2.5$.
For $\beta=4$, $Y_2$ displays  a minimum similarly to the C2C model, but,
as shown in the inset, there is an important difference between 
the two models because
$\gamma$(C2C)$= 3$, while $\gamma(\beta=4) = 2$.
Accordingly, for finite systems, delocalized states extend above the critical threshold up to  sizes of
order $N^*$, with
\be
N^*(\beta =4) \sim \delta^{-2} \ll N^*(C2C) \sim \delta^{-3}. 
\ee 
From the derivation of the curves $K(N,\delta)$ as explained for the C2C model and from  Eq.~(\ref{eq.Y2anal}), again,
we are able to reproduce the $N-$dependence of the participation ratio for different values of $\delta$,
see Fig.~\ref{fig.beta4} (main). 

When the exponent $\beta$ is reduced, the minimum of $Y_2$ can even disappear, as observed for $\beta=2.5$, 
see Fig.~\ref{fig.beta2.5}. The monotonic decrease of $Y_2$ towards the finite asymptotic value
is compatible with a localization transition involving condensation on a single site also for finite $N$. 
In fact, for $\beta=2.5$ no peak exists in $K$ versus $N$, see the inset. The absence of a decreasing
regime for $K$ implies the absence of a minimum for $Y_2$.
Furthermore, we observe that the first term on the RHS of Eq.~(\ref{eq.Y2anal}) does not decrease
 as $1/N$ because the second moment  $\mu_2$ at the numerator diverges for $\beta<3$. 
More precisely, such term vanishes as $1/N^{\beta -2}$, which explains the decay $Y_2 \sim 1/\sqrt{N}$ 
observed in Fig.~\ref{fig.beta2.5}.
In the absence of a minimum, $N^*$ can be interpreted as a size separating the delocalized phase 
from the localized one and it can be derived by equating the two terms of Eq.~(\ref{eq.Y2anal}).
For $\beta=2.5$ we have $1/\sqrt{N^*} \sim \delta^2$, therefore $N^*(\beta=2.5) = \delta^{-4}$.
Using the relation $N^*(\beta) \simeq \delta^{-\gamma}$ we can define 
the exponent $\gamma$ also if $Y_2$ does not display a minimum and obtain $\gamma(\beta=2.5)=4$.

\section{Discussion}
\label{sec.discussion}

In the thermodynamic limit the main properties of the localization transition are well understood:
(i)~Eq.~(\ref{eq.condcond}) gives the condensation condition for the class of the discrete 
Zero Range Process (ZRP) models
and for their continuous counterpart, that we called here Factorized Continuous Transport (FCT) models.
(ii)~At criticality the single-site marginal distribution function $p(\e)$ in Eq.~(\ref{eq.dressed}) equals the function $f(\e)$.
Above criticality $p(\e)$ is given by $f(\e)$
plus a Dirac delta centred at the energy excess $N\delta$;
(iii)~The value of the participation ratio in the thermodynamic limit is simply $Y_2^\infty(\delta)=(\delta/h)^2$.

When the finite size $N$ is taken into account things get more complicated but also more interesting,
as physically relevant phenomena may appear.
In this paper, we studied numerically several condensation  models 
focusing on the behavior of the participation ratio $Y_2(N,\delta)$, which is the standard order parameter~\cite{Godreche2012_JSTAT,Gradenigo2017_Entropy,Gradenigo2021_JSTAT} as well as an
experimentally accessible observable, see e.g.~\cite{lahini09}.

Concerning the C2C model, our results are in agreement with the recent analytical results
obtained in Ref.~\cite{Gradenigo2021_JSTAT} using the large deviation theory.
Our study confirms that for finite $N$ the system is still nonlocalized for $\delta < \delta^* \sim 1/N^{1/3}$
and it shows that
 the transition between the nonlocalized and localized regimes corresponds
to a minimum of the participation ratio. The large extension of the nonlocalized region
is related to an exponent $\gamma=3$. 
In the context of the DNLS equation where localized states are  generated dynamically as {\it discrete breather} 
excitations~\cite{flach98,flach08},
these results corroborate the idea that multi-breather states are compatible with equilibrium conditions, 
 as  discussed in~\cite{Iubini2013_NJP}. 
We have also shown that, consistently with~\cite{Gradenigo2021_JSTAT}, analogous results hold for a canonical
FCT model characterized by a stretched-exponential distribution. 

The above properties are modified for FCT models 
with power-law distributions,
$f(\e) \sim 1/\e^\beta$ $(\beta>2)$.
More precisely, we have studied two cases, $\beta=4$ and $\beta=2.5$.
For $\beta=4$, $Y_2$ still displays a minimum in $N$, yet  $\gamma=2$. This means that $\gamma$ has the lowest possible
value, so that, for finite $N$ and $\delta>0$,  the delocalized region has the smallest possible extension in $N$.
For $\beta=2.5$, $Y_2$ does not even have a minimum. Moreover, since the second moment of the distribution
diverges, the nonlocalized region is ``strongly" nonlocalized, because $Y_2$ decays as $1/\sqrt{N}$
rather than as $1/N$. Even if a minimum does not exist, it is possible to define $N^*$ as the length separating
the nonlocalized region $(N \ll N^*)$ from the localized one $(N\gg N^*)$ and it is therefore possible to define 
the exponent $\gamma$ through the relation $N^* \simeq \delta^{-\gamma}$. With this definition, we can sum up
our findings: $\gamma(\beta=2.5) = 4$, $\gamma(\beta=4) = 2$, $\gamma(\mbox{stretched-exponential}) = \gamma(\mbox{C2C}) = 3$.

We also stress that significantly different localization scenarios can have a unified interpretation 
through the effective number of sites hosting the condensate, $K(N,\delta)$, which
determines the participation ratio, see Eq.~(\ref{eq.Y2anal}). In fact, the minimum of $Y_2(N)$
is related to a maximum of $K(N)$ and the exponent $\gamma >2$ is related to a maximum of $K(N)$ whose value increases
upon decreasing $\delta$. The non-monotonic behavior of $Y_2(N)$ has also important physical 
implications for what concerns the observation of effectively delocalized states above the critical condensation point,
as recently discussed in~\cite{Gradenigo2021_JSTAT} for the DNLS model.

\begin{figure}
\begin{center}
\includegraphics[width=0.49\textwidth,clip]{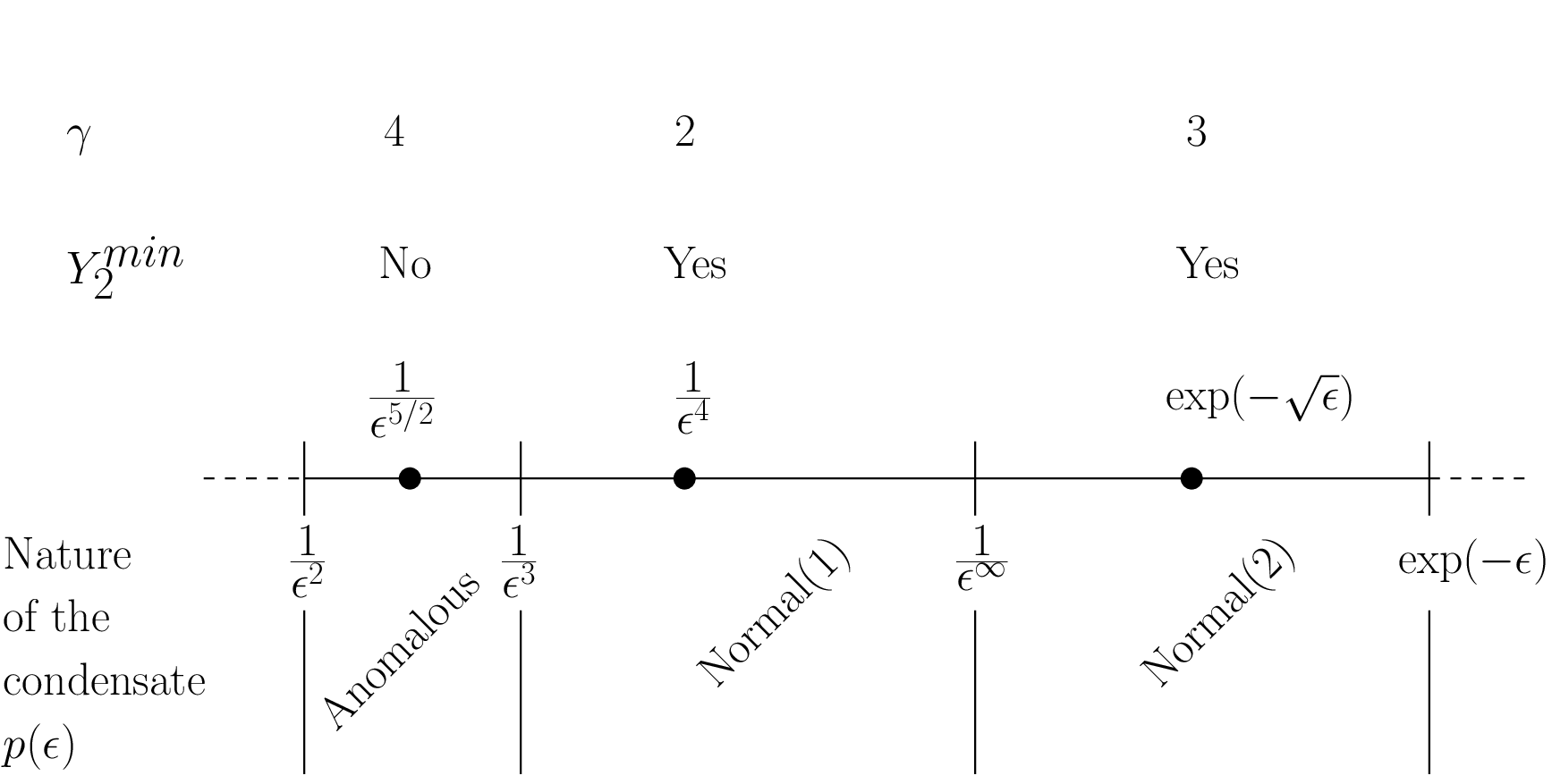}
\end{center}
\caption{
Localization properties for canonical FCT models with distributions in the range $1/\e^2 > f(\e) >\exp(-\e) $.
Upper half: behavior of $Y_2^{min}$ and $\gamma$ obtained in the present article for the specific models
indicated by the full dots (the results for the stretched exponential case are
also representative of the C2C model).
Lower half: nature of the condensate as obtained in~\cite{Evans2006_JSP} through the single-site marginal 
distribution $p(\e)$.
Normal (Anomalous) means that fluctuations of the condensate are (are not) gaussian. 
Normal$(1)$ and Normal$(2)$ refer to different scales over which fluctuations are gaussian (see the main text).
}
\label{fig.diagram}
\end{figure}

It is useful to compare our results with related 
studies of the nature of the condensate based on the single-site marginal distribution~\cite{Evans2006_JSP}.
The case $\beta=4$ corresponds to gaussian fluctuations of the condensate (called normal condensate)
on the scale $|\e-\delta N|\sim N^{2/3}$
while the case $\beta=2.5$ corresponds to non gaussian fluctuations of the condensate (called anomalous condensate).
Finally, the stretched-exponential case is characterized by gaussian fluctuations but on the different scale
$|\e-\delta N|\sim N^{1/2}$.
Additionally, in~\cite{Evans2006_JSP} it was also noted that the nature of the condensation transition  for stretched exponential
distributions
differs from the usual scenario of a second order phase transition. In fact, specific signatures of a first order
phase transition have been recently found in~\cite{gradenigo19,Gradenigo2021_JSTAT}.
In Fig.~\ref{fig.diagram} we show a schematic phase diagram which summarizes our results and the results obtained 
in~\cite{Evans2006_JSP}. We also mention that we have obtained preliminary results for the model $\beta=6$ and
for a modified stretched-exponential distribution, $f(\epsilon)\sim \exp(-\epsilon^{0.6})$.
Results clearly show that in both cases $Y_2$ has a minimum in $N$ and suggest that
$\gamma=2$ for $\beta=6$ and $\gamma=3$ for the modified stretched-exponential case.

In order to clarify the full localization scenarios in the whole range of distributions given in Eq.~(\ref{eq.condcond}) 
it will be necessary to perform
detailed simulations for several other models, keeping in mind that larger values of $\gamma$
would be very problematic for numerics, because $N^*$ scales as $1/\delta^\gamma$,
therefore forcing to investigate extremely large systems.
Alternatively, analytic approaches allowing to extract the peculiar features
that make a model fall in a given scenario would be desirable.
We hope that our results might lead to pursue further analytical studies to find
a rigorous connection between the properties of $Y_2$ and $K$ on the one hand and those of $p(\e)$ on the other hand.

\begin{acknowledgments}
Authors thank Onofrio Mazzarisi for data concerning models where $\e_i$ is an integer variable.
They also thank Federico Corberi for several discussions and Antonio Politi for comments on the preprint.
PP acknowledges support from the MIUR PRIN 2017 project 201798CZLJ.
\end{acknowledgments}


\begin{thebibliography}{36}%
\makeatletter
\providecommand \@ifxundefined [1]{%
 \@ifx{#1\undefined}
}%
\providecommand \@ifnum [1]{%
 \ifnum #1\expandafter \@firstoftwo
 \else \expandafter \@secondoftwo
 \fi
}%
\providecommand \@ifx [1]{%
 \ifx #1\expandafter \@firstoftwo
 \else \expandafter \@secondoftwo
 \fi
}%
\providecommand \natexlab [1]{#1}%
\providecommand \enquote  [1]{``#1''}%
\providecommand \bibnamefont  [1]{#1}%
\providecommand \bibfnamefont [1]{#1}%
\providecommand \citenamefont [1]{#1}%
\providecommand \href@noop [0]{\@secondoftwo}%
\providecommand \href [0]{\begingroup \@sanitize@url \@href}%
\providecommand \@href[1]{\@@startlink{#1}\@@href}%
\providecommand \@@href[1]{\endgroup#1\@@endlink}%
\providecommand \@sanitize@url [0]{\catcode `\\12\catcode `\$12\catcode
  `\&12\catcode `\#12\catcode `\^12\catcode `\_12\catcode `\%12\relax}%
\providecommand \@@startlink[1]{}%
\providecommand \@@endlink[0]{}%
\providecommand \url  [0]{\begingroup\@sanitize@url \@url }%
\providecommand \@url [1]{\endgroup\@href {#1}{\urlprefix }}%
\providecommand \urlprefix  [0]{URL }%
\providecommand \Eprint [0]{\href }%
\providecommand \doibase [0]{http://dx.doi.org/}%
\providecommand \selectlanguage [0]{\@gobble}%
\providecommand \bibinfo  [0]{\@secondoftwo}%
\providecommand \bibfield  [0]{\@secondoftwo}%
\providecommand \translation [1]{[#1]}%
\providecommand \BibitemOpen [0]{}%
\providecommand \bibitemStop [0]{}%
\providecommand \bibitemNoStop [0]{.\EOS\space}%
\providecommand \EOS [0]{\spacefactor3000\relax}%
\providecommand \BibitemShut  [1]{\csname bibitem#1\endcsname}%
\let\auto@bib@innerbib\@empty
\bibitem [{\citenamefont {Chaikin}\ \emph {et~al.}(1995)\citenamefont
  {Chaikin}, \citenamefont {Lubensky},\ and\ \citenamefont
  {Witten}}]{book_Chaikin}%
  \BibitemOpen
  \bibfield  {author} {\bibinfo {author} {\bibfnamefont {P.~M.}\ \bibnamefont
  {Chaikin}}, \bibinfo {author} {\bibfnamefont {T.~C.}\ \bibnamefont
  {Lubensky}}, \ and\ \bibinfo {author} {\bibfnamefont {T.~A.}\ \bibnamefont
  {Witten}},\ }\href@noop {} {\emph {\bibinfo {title} {Principles of condensed
  matter physics}}},\ Vol.~\bibinfo {volume} {10}\ (\bibinfo  {publisher}
  {Cambridge university press Cambridge},\ \bibinfo {year} {1995})\BibitemShut
  {NoStop}%
\bibitem [{\citenamefont {Huang}(1987)}]{book_Huang}%
  \BibitemOpen
  \bibfield  {author} {\bibinfo {author} {\bibfnamefont {K.}~\bibnamefont
  {Huang}},\ }\href@noop {} {\enquote {\bibinfo {title} {Statistical mechanics
  2nd edn},}\ } (\bibinfo {year} {1987})\BibitemShut {NoStop}%
\bibitem [{\citenamefont {Drouffe}\ \emph {et~al.}(1998)\citenamefont
  {Drouffe}, \citenamefont {Godr{\`e}che},\ and\ \citenamefont
  {Camia}}]{Drouffe1998_JPA}%
  \BibitemOpen
  \bibfield  {author} {\bibinfo {author} {\bibfnamefont {J.-M.}\ \bibnamefont
  {Drouffe}}, \bibinfo {author} {\bibfnamefont {C.}~\bibnamefont
  {Godr{\`e}che}}, \ and\ \bibinfo {author} {\bibfnamefont {F.}~\bibnamefont
  {Camia}},\ }\href@noop {} {\bibfield  {journal} {\bibinfo  {journal} {Journal
  of Physics A: Mathematical and General}\ }\textbf {\bibinfo {volume} {31}},\
  \bibinfo {pages} {L19} (\bibinfo {year} {1998})}\BibitemShut {NoStop}%
\bibitem [{\citenamefont {Majumdar}\ \emph {et~al.}(2005)\citenamefont
  {Majumdar}, \citenamefont {Evans},\ and\ \citenamefont
  {Zia}}]{Majumdar2005_PRL}%
  \BibitemOpen
  \bibfield  {author} {\bibinfo {author} {\bibfnamefont {S.~N.}\ \bibnamefont
  {Majumdar}}, \bibinfo {author} {\bibfnamefont {M.}~\bibnamefont {Evans}}, \
  and\ \bibinfo {author} {\bibfnamefont {R.}~\bibnamefont {Zia}},\ }\href@noop
  {} {\bibfield  {journal} {\bibinfo  {journal} {Physical review letters}\
  }\textbf {\bibinfo {volume} {94}},\ \bibinfo {pages} {180601} (\bibinfo
  {year} {2005})}\BibitemShut {NoStop}%
\bibitem [{\citenamefont {Godreche}(2007)}]{Godreche2007_LNP}%
  \BibitemOpen
  \bibfield  {author} {\bibinfo {author} {\bibfnamefont {C.}~\bibnamefont
  {Godreche}},\ }in\ \href@noop {} {\emph {\bibinfo {booktitle} {Ageing and the
  glass transition}}}\ (\bibinfo  {publisher} {Springer},\ \bibinfo {year}
  {2007})\ pp.\ \bibinfo {pages} {261--294}\BibitemShut {NoStop}%
\bibitem [{\citenamefont {Majumdar}(2010)}]{Majumdar2010_LesHouches}%
  \BibitemOpen
  \bibfield  {author} {\bibinfo {author} {\bibfnamefont {S.}~\bibnamefont
  {Majumdar}},\ }\href@noop {} {\bibfield  {journal} {\bibinfo  {journal}
  {Exact Methods in Low-dimensional Statistical Physics and Quantum Computing:
  Lecture Notes of the Les Houches Summer School: Volume 89, July 2008}\ ,\
  \bibinfo {pages} {407}} (\bibinfo {year} {2010})}\BibitemShut {NoStop}%
\bibitem [{\citenamefont {Zannetti}\ \emph {et~al.}(2014)\citenamefont
  {Zannetti}, \citenamefont {Corberi},\ and\ \citenamefont
  {Gonnella}}]{Zannetti2014_PRE}%
  \BibitemOpen
  \bibfield  {author} {\bibinfo {author} {\bibfnamefont {M.}~\bibnamefont
  {Zannetti}}, \bibinfo {author} {\bibfnamefont {F.}~\bibnamefont {Corberi}}, \
  and\ \bibinfo {author} {\bibfnamefont {G.}~\bibnamefont {Gonnella}},\
  }\href@noop {} {\bibfield  {journal} {\bibinfo  {journal} {Physical Review
  E}\ }\textbf {\bibinfo {volume} {90}},\ \bibinfo {pages} {012143} (\bibinfo
  {year} {2014})}\BibitemShut {NoStop}%
\bibitem [{\citenamefont {Godr{\`e}che}(2020)}]{Godreche2020_arXiv}%
  \BibitemOpen
  \bibfield  {author} {\bibinfo {author} {\bibfnamefont {C.}~\bibnamefont
  {Godr{\`e}che}},\ }\href@noop {} {\bibfield  {journal} {\bibinfo  {journal}
  {arXiv preprint arXiv:2006.04076}\ } (\bibinfo {year} {2020})}\BibitemShut
  {NoStop}%
\bibitem [{\citenamefont {Eggers}(1999)}]{eggers99}%
  \BibitemOpen
  \bibfield  {author} {\bibinfo {author} {\bibfnamefont {J.}~\bibnamefont
  {Eggers}},\ }\href@noop {} {\bibfield  {journal} {\bibinfo  {journal}
  {Physical Review Letters}\ }\textbf {\bibinfo {volume} {83}},\ \bibinfo
  {pages} {5322} (\bibinfo {year} {1999})}\BibitemShut {NoStop}%
\bibitem [{\citenamefont {T{\"o}r{\"o}k}(2005)}]{torok05}%
  \BibitemOpen
  \bibfield  {author} {\bibinfo {author} {\bibfnamefont {J.}~\bibnamefont
  {T{\"o}r{\"o}k}},\ }\href@noop {} {\bibfield  {journal} {\bibinfo  {journal}
  {Physica A: Statistical Mechanics and its Applications}\ }\textbf {\bibinfo
  {volume} {355}},\ \bibinfo {pages} {374} (\bibinfo {year}
  {2005})}\BibitemShut {NoStop}%
\bibitem [{\citenamefont {Evans}\ and\ \citenamefont
  {Hanney}(2005)}]{Evans2005_JPA}%
  \BibitemOpen
  \bibfield  {author} {\bibinfo {author} {\bibfnamefont {M.~R.}\ \bibnamefont
  {Evans}}\ and\ \bibinfo {author} {\bibfnamefont {T.}~\bibnamefont {Hanney}},\
  }\href@noop {} {\bibfield  {journal} {\bibinfo  {journal} {Journal of Physics
  A: Mathematical and General}\ }\textbf {\bibinfo {volume} {38}},\ \bibinfo
  {pages} {R195} (\bibinfo {year} {2005})}\BibitemShut {NoStop}%
\bibitem [{\citenamefont {Kevrekidis}(2009)}]{kevrekidis09}%
  \BibitemOpen
  \bibfield  {author} {\bibinfo {author} {\bibfnamefont {P.~G.}\ \bibnamefont
  {Kevrekidis}},\ }\href@noop {} {\emph {\bibinfo {title} {The discrete
  nonlinear Schr{\"o}dinger equation: mathematical analysis, numerical
  computations and physical perspectives}}},\ Vol.\ \bibinfo {volume} {232}\
  (\bibinfo  {publisher} {Springer Science \& Business Media},\ \bibinfo {year}
  {2009})\BibitemShut {NoStop}%
\bibitem [{\citenamefont {Eisenberg}\ \emph {et~al.}(1998)\citenamefont
  {Eisenberg}, \citenamefont {Silberberg}, \citenamefont {Morandotti},
  \citenamefont {Boyd},\ and\ \citenamefont {Aitchison}}]{ESMBA}%
  \BibitemOpen
  \bibfield  {author} {\bibinfo {author} {\bibfnamefont {H.}~\bibnamefont
  {Eisenberg}}, \bibinfo {author} {\bibfnamefont {Y.}~\bibnamefont
  {Silberberg}}, \bibinfo {author} {\bibfnamefont {R.}~\bibnamefont
  {Morandotti}}, \bibinfo {author} {\bibfnamefont {A.}~\bibnamefont {Boyd}}, \
  and\ \bibinfo {author} {\bibfnamefont {J.}~\bibnamefont {Aitchison}},\
  }\href@noop {} {\bibfield  {journal} {\bibinfo  {journal} {Physical Review
  Letters}\ }\textbf {\bibinfo {volume} {81}},\ \bibinfo {pages} {3383}
  (\bibinfo {year} {1998})}\BibitemShut {NoStop}%
\bibitem [{\citenamefont {Trombettoni}\ and\ \citenamefont
  {Smerzi}(2001)}]{TS}%
  \BibitemOpen
  \bibfield  {author} {\bibinfo {author} {\bibfnamefont {A.}~\bibnamefont
  {Trombettoni}}\ and\ \bibinfo {author} {\bibfnamefont {A.}~\bibnamefont
  {Smerzi}},\ }\href@noop {} {\bibfield  {journal} {\bibinfo  {journal}
  {Physical Review Letters}\ }\textbf {\bibinfo {volume} {86}},\ \bibinfo
  {pages} {2353} (\bibinfo {year} {2001})}\BibitemShut {NoStop}%
\bibitem [{\citenamefont {Gradenigo}\ \emph
  {et~al.}(2021{\natexlab{a}})\citenamefont {Gradenigo}, \citenamefont
  {Iubini}, \citenamefont {Livi},\ and\ \citenamefont
  {Majumdar}}]{Gradenigo2021_JSTAT}%
  \BibitemOpen
  \bibfield  {author} {\bibinfo {author} {\bibfnamefont {G.}~\bibnamefont
  {Gradenigo}}, \bibinfo {author} {\bibfnamefont {S.}~\bibnamefont {Iubini}},
  \bibinfo {author} {\bibfnamefont {R.}~\bibnamefont {Livi}}, \ and\ \bibinfo
  {author} {\bibfnamefont {S.~N.}\ \bibnamefont {Majumdar}},\ }\href {\doibase
  10.1088/1742-5468/abda26} {\bibfield  {journal} {\bibinfo  {journal} {Journal
  of Statistical Mechanics: Theory and Experiment}\ }\textbf {\bibinfo {volume}
  {2021}},\ \bibinfo {pages} {023201} (\bibinfo {year}
  {2021}{\natexlab{a}})}\BibitemShut {NoStop}%
\bibitem [{\citenamefont {Gradenigo}\ \emph
  {et~al.}(2021{\natexlab{b}})\citenamefont {Gradenigo}, \citenamefont
  {Iubini}, \citenamefont {Livi},\ and\ \citenamefont
  {Majumdar}}]{Gradenigo2021_EPJE}%
  \BibitemOpen
  \bibfield  {author} {\bibinfo {author} {\bibfnamefont {G.}~\bibnamefont
  {Gradenigo}}, \bibinfo {author} {\bibfnamefont {S.}~\bibnamefont {Iubini}},
  \bibinfo {author} {\bibfnamefont {R.}~\bibnamefont {Livi}}, \ and\ \bibinfo
  {author} {\bibfnamefont {S.~N.}\ \bibnamefont {Majumdar}},\ }\href@noop {}
  {\bibfield  {journal} {\bibinfo  {journal} {The European Physical Journal E}\
  }\textbf {\bibinfo {volume} {44}},\ \bibinfo {pages} {1} (\bibinfo {year}
  {2021}{\natexlab{b}})}\BibitemShut {NoStop}%
\bibitem [{\citenamefont {Evans}\ \emph {et~al.}(2004)\citenamefont {Evans},
  \citenamefont {Majumdar},\ and\ \citenamefont {Zia}}]{evans04jpa}%
  \BibitemOpen
  \bibfield  {author} {\bibinfo {author} {\bibfnamefont {M.~R.}\ \bibnamefont
  {Evans}}, \bibinfo {author} {\bibfnamefont {S.~N.}\ \bibnamefont {Majumdar}},
  \ and\ \bibinfo {author} {\bibfnamefont {R.~K.}\ \bibnamefont {Zia}},\
  }\href@noop {} {\bibfield  {journal} {\bibinfo  {journal} {Journal of Physics
  A: Mathematical and General}\ }\textbf {\bibinfo {volume} {37}},\ \bibinfo
  {pages} {L275} (\bibinfo {year} {2004})}\BibitemShut {NoStop}%
\bibitem [{\citenamefont {Zia}\ \emph {et~al.}(2004)\citenamefont {Zia},
  \citenamefont {Evans},\ and\ \citenamefont {Majumdar}}]{zia04}%
  \BibitemOpen
  \bibfield  {author} {\bibinfo {author} {\bibfnamefont {R.}~\bibnamefont
  {Zia}}, \bibinfo {author} {\bibfnamefont {M.}~\bibnamefont {Evans}}, \ and\
  \bibinfo {author} {\bibfnamefont {S.~N.}\ \bibnamefont {Majumdar}},\
  }\href@noop {} {\bibfield  {journal} {\bibinfo  {journal} {Journal of
  Statistical Mechanics: Theory and Experiment}\ }\textbf {\bibinfo {volume}
  {2004}},\ \bibinfo {pages} {L10001} (\bibinfo {year} {2004})}\BibitemShut
  {NoStop}%
\bibitem [{\citenamefont {Evans}\ \emph {et~al.}(2006)\citenamefont {Evans},
  \citenamefont {Majumdar},\ and\ \citenamefont {Zia}}]{Evans2006_JSP}%
  \BibitemOpen
  \bibfield  {author} {\bibinfo {author} {\bibfnamefont {M.}~\bibnamefont
  {Evans}}, \bibinfo {author} {\bibfnamefont {S.~N.}\ \bibnamefont {Majumdar}},
  \ and\ \bibinfo {author} {\bibfnamefont {R.}~\bibnamefont {Zia}},\
  }\href@noop {} {\bibfield  {journal} {\bibinfo  {journal} {Journal of
  Statistical Physics}\ }\textbf {\bibinfo {volume} {123}},\ \bibinfo {pages}
  {357} (\bibinfo {year} {2006})}\BibitemShut {NoStop}%
\bibitem [{\citenamefont {Godr{\`e}che}\ and\ \citenamefont
  {Luck}(2001)}]{Godreche2001_EPJB}%
  \BibitemOpen
  \bibfield  {author} {\bibinfo {author} {\bibfnamefont {C.}~\bibnamefont
  {Godr{\`e}che}}\ and\ \bibinfo {author} {\bibfnamefont {J.}~\bibnamefont
  {Luck}},\ }\href@noop {} {\bibfield  {journal} {\bibinfo  {journal} {The
  European Physical Journal B-Condensed Matter and Complex Systems}\ }\textbf
  {\bibinfo {volume} {23}},\ \bibinfo {pages} {473} (\bibinfo {year}
  {2001})}\BibitemShut {NoStop}%
\bibitem [{\citenamefont {Thouless}(1974)}]{Thouless1974_review}%
  \BibitemOpen
  \bibfield  {author} {\bibinfo {author} {\bibfnamefont {D.~J.}\ \bibnamefont
  {Thouless}},\ }\href@noop {} {\bibfield  {journal} {\bibinfo  {journal}
  {Physics Reports}\ }\textbf {\bibinfo {volume} {13}},\ \bibinfo {pages} {93}
  (\bibinfo {year} {1974})}\BibitemShut {NoStop}%
\bibitem [{\citenamefont {Gradenigo}\ and\ \citenamefont
  {Bertin}(2017)}]{Gradenigo2017_Entropy}%
  \BibitemOpen
  \bibfield  {author} {\bibinfo {author} {\bibfnamefont {G.}~\bibnamefont
  {Gradenigo}}\ and\ \bibinfo {author} {\bibfnamefont {E.}~\bibnamefont
  {Bertin}},\ }\href@noop {} {\bibfield  {journal} {\bibinfo  {journal}
  {Entropy}\ }\textbf {\bibinfo {volume} {19}},\ \bibinfo {pages} {517}
  (\bibinfo {year} {2017})}\BibitemShut {NoStop}%
\bibitem [{\citenamefont {Rasmussen}\ \emph {et~al.}(2000)\citenamefont
  {Rasmussen}, \citenamefont {Cretegny}, \citenamefont {Kevrekidis},\ and\
  \citenamefont {Gr{\o}nbech-Jensen}}]{Rasmussen2000_PRL}%
  \BibitemOpen
  \bibfield  {author} {\bibinfo {author} {\bibfnamefont {K.}~\bibnamefont
  {Rasmussen}}, \bibinfo {author} {\bibfnamefont {T.}~\bibnamefont {Cretegny}},
  \bibinfo {author} {\bibfnamefont {P.~G.}\ \bibnamefont {Kevrekidis}}, \ and\
  \bibinfo {author} {\bibfnamefont {N.}~\bibnamefont {Gr{\o}nbech-Jensen}},\
  }\href@noop {} {\bibfield  {journal} {\bibinfo  {journal} {Physical review
  letters}\ }\textbf {\bibinfo {volume} {84}},\ \bibinfo {pages} {3740}
  (\bibinfo {year} {2000})}\BibitemShut {NoStop}%
\bibitem [{\citenamefont {Iubini}\ \emph {et~al.}(2013)\citenamefont {Iubini},
  \citenamefont {Franzosi}, \citenamefont {Livi}, \citenamefont {Oppo},\ and\
  \citenamefont {Politi}}]{Iubini2013_NJP}%
  \BibitemOpen
  \bibfield  {author} {\bibinfo {author} {\bibfnamefont {S.}~\bibnamefont
  {Iubini}}, \bibinfo {author} {\bibfnamefont {R.}~\bibnamefont {Franzosi}},
  \bibinfo {author} {\bibfnamefont {R.}~\bibnamefont {Livi}}, \bibinfo {author}
  {\bibfnamefont {G.-L.}\ \bibnamefont {Oppo}}, \ and\ \bibinfo {author}
  {\bibfnamefont {A.}~\bibnamefont {Politi}},\ }\href@noop {} {\bibfield
  {journal} {\bibinfo  {journal} {New Journal of Physics}\ }\textbf {\bibinfo
  {volume} {15}},\ \bibinfo {pages} {023032} (\bibinfo {year}
  {2013})}\BibitemShut {NoStop}%
\bibitem [{\citenamefont {Iubini}\ \emph {et~al.}(2014)\citenamefont {Iubini},
  \citenamefont {Politi},\ and\ \citenamefont {Politi}}]{JSP_DNLS}%
  \BibitemOpen
  \bibfield  {author} {\bibinfo {author} {\bibfnamefont {S.}~\bibnamefont
  {Iubini}}, \bibinfo {author} {\bibfnamefont {A.}~\bibnamefont {Politi}}, \
  and\ \bibinfo {author} {\bibfnamefont {P.}~\bibnamefont {Politi}},\
  }\href@noop {} {\bibfield  {journal} {\bibinfo  {journal} {Journal of
  Statistical Physics}\ }\textbf {\bibinfo {volume} {154}},\ \bibinfo {pages}
  {1057} (\bibinfo {year} {2014})}\BibitemShut {NoStop}%
\bibitem [{\citenamefont {Szavits-Nossan}\ \emph {et~al.}(2014)\citenamefont
  {Szavits-Nossan}, \citenamefont {Evans},\ and\ \citenamefont
  {Majumdar}}]{Szavits2014_PRL}%
  \BibitemOpen
  \bibfield  {author} {\bibinfo {author} {\bibfnamefont {J.}~\bibnamefont
  {Szavits-Nossan}}, \bibinfo {author} {\bibfnamefont {M.~R.}\ \bibnamefont
  {Evans}}, \ and\ \bibinfo {author} {\bibfnamefont {S.~N.}\ \bibnamefont
  {Majumdar}},\ }\href@noop {} {\bibfield  {journal} {\bibinfo  {journal}
  {Physical review letters}\ }\textbf {\bibinfo {volume} {112}},\ \bibinfo
  {pages} {020602} (\bibinfo {year} {2014})}\BibitemShut {NoStop}%
\bibitem [{\citenamefont {Iubini}\ \emph {et~al.}(2017)\citenamefont {Iubini},
  \citenamefont {Politi},\ and\ \citenamefont {Politi}}]{JSTAT_mmc}%
  \BibitemOpen
  \bibfield  {author} {\bibinfo {author} {\bibfnamefont {S.}~\bibnamefont
  {Iubini}}, \bibinfo {author} {\bibfnamefont {A.}~\bibnamefont {Politi}}, \
  and\ \bibinfo {author} {\bibfnamefont {P.}~\bibnamefont {Politi}},\
  }\href@noop {} {\bibfield  {journal} {\bibinfo  {journal} {Journal of
  Statistical Mechanics: Theory and Experiment}\ }\textbf {\bibinfo {volume}
  {2017}},\ \bibinfo {pages} {073201} (\bibinfo {year} {2017})}\BibitemShut
  {NoStop}%
\bibitem [{Note1()}]{Note1}%
  \BibitemOpen
  \bibinfo {note} {We have always verified for all numerical simulations that
  the equilibrium state was reached: suitable relaxation transients were
  introduced to make the system relax before computing statistical
  averages.}\BibitemShut {Stop}%
\bibitem [{Note2()}]{Note2}%
  \BibitemOpen
  \bibinfo {note} {Eq. (\ref {eq.C2Cstretched}) corresponds to a Weibull
  distribution~\cite {Weibull} in the variable $\epsilon $ with shape parameter
  $1/2$ and scale parameter $\lambda ^{-2}$.}\BibitemShut {Stop}%
\bibitem [{Note3()}]{Note3}%
  \BibitemOpen
  \bibinfo {note} {This is because for a system of size $N$ the maximal
  possible energy on a site is $Nh$, not infinity.}\BibitemShut {Stop}%
\bibitem [{\citenamefont {Godr{\`e}che}\ and\ \citenamefont
  {Luck}(2012)}]{Godreche2012_JSTAT}%
  \BibitemOpen
  \bibfield  {author} {\bibinfo {author} {\bibfnamefont {C.}~\bibnamefont
  {Godr{\`e}che}}\ and\ \bibinfo {author} {\bibfnamefont {J.-M.}\ \bibnamefont
  {Luck}},\ }\href@noop {} {\bibfield  {journal} {\bibinfo  {journal} {Journal
  of Statistical Mechanics: Theory and Experiment}\ }\textbf {\bibinfo {volume}
  {2012}},\ \bibinfo {pages} {P12013} (\bibinfo {year} {2012})}\BibitemShut
  {NoStop}%
\bibitem [{\citenamefont {Lahini}\ \emph {et~al.}(2009)\citenamefont {Lahini},
  \citenamefont {Pugatch}, \citenamefont {Pozzi}, \citenamefont {Sorel},
  \citenamefont {Morandotti}, \citenamefont {Davidson},\ and\ \citenamefont
  {Silberberg}}]{lahini09}%
  \BibitemOpen
  \bibfield  {author} {\bibinfo {author} {\bibfnamefont {Y.}~\bibnamefont
  {Lahini}}, \bibinfo {author} {\bibfnamefont {R.}~\bibnamefont {Pugatch}},
  \bibinfo {author} {\bibfnamefont {F.}~\bibnamefont {Pozzi}}, \bibinfo
  {author} {\bibfnamefont {M.}~\bibnamefont {Sorel}}, \bibinfo {author}
  {\bibfnamefont {R.}~\bibnamefont {Morandotti}}, \bibinfo {author}
  {\bibfnamefont {N.}~\bibnamefont {Davidson}}, \ and\ \bibinfo {author}
  {\bibfnamefont {Y.}~\bibnamefont {Silberberg}},\ }\href@noop {} {\bibfield
  {journal} {\bibinfo  {journal} {Physical Review Letters}\ }\textbf {\bibinfo
  {volume} {103}},\ \bibinfo {pages} {013901} (\bibinfo {year}
  {2009})}\BibitemShut {NoStop}%
\bibitem [{\citenamefont {Flach}\ and\ \citenamefont {Willis}(1998)}]{flach98}%
  \BibitemOpen
  \bibfield  {author} {\bibinfo {author} {\bibfnamefont {S.}~\bibnamefont
  {Flach}}\ and\ \bibinfo {author} {\bibfnamefont {C.~R.}\ \bibnamefont
  {Willis}},\ }\href@noop {} {\bibfield  {journal} {\bibinfo  {journal}
  {Physics Reports}\ }\textbf {\bibinfo {volume} {295}},\ \bibinfo {pages}
  {181} (\bibinfo {year} {1998})}\BibitemShut {NoStop}%
\bibitem [{\citenamefont {Flach}\ and\ \citenamefont
  {Gorbach}(2008)}]{flach08}%
  \BibitemOpen
  \bibfield  {author} {\bibinfo {author} {\bibfnamefont {S.}~\bibnamefont
  {Flach}}\ and\ \bibinfo {author} {\bibfnamefont {A.~V.}\ \bibnamefont
  {Gorbach}},\ }\href@noop {} {\bibfield  {journal} {\bibinfo  {journal}
  {Physics Reports}\ }\textbf {\bibinfo {volume} {467}},\ \bibinfo {pages} {1}
  (\bibinfo {year} {2008})}\BibitemShut {NoStop}%
\bibitem [{\citenamefont {Gradenigo}\ and\ \citenamefont
  {Majumdar}(2019)}]{gradenigo19}%
  \BibitemOpen
  \bibfield  {author} {\bibinfo {author} {\bibfnamefont {G.}~\bibnamefont
  {Gradenigo}}\ and\ \bibinfo {author} {\bibfnamefont {S.~N.}\ \bibnamefont
  {Majumdar}},\ }\href@noop {} {\bibfield  {journal} {\bibinfo  {journal}
  {Journal of Statistical Mechanics: Theory and Experiment}\ }\textbf {\bibinfo
  {volume} {2019}},\ \bibinfo {pages} {053206} (\bibinfo {year}
  {2019})}\BibitemShut {NoStop}%
\bibitem [{\citenamefont {Forbes}\ \emph {et~al.}(2011)\citenamefont {Forbes},
  \citenamefont {Evans}, \citenamefont {Hastings},\ and\ \citenamefont
  {Peacock}}]{Weibull}%
  \BibitemOpen
  \bibfield  {author} {\bibinfo {author} {\bibfnamefont {C.}~\bibnamefont
  {Forbes}}, \bibinfo {author} {\bibfnamefont {M.}~\bibnamefont {Evans}},
  \bibinfo {author} {\bibfnamefont {N.}~\bibnamefont {Hastings}}, \ and\
  \bibinfo {author} {\bibfnamefont {B.}~\bibnamefont {Peacock}},\ }\href@noop
  {} {\emph {\bibinfo {title} {Statistical distributions}}}\ (\bibinfo
  {publisher} {John Wiley \& Sons},\ \bibinfo {year} {2011})\BibitemShut
  {NoStop}%
\end{thebibliography}

%

\end{document}